\documentclass[twocolumn]{aastex631}
\usepackage{graphicx,times}             
\usepackage{color}
\usepackage{ulem}

\newcommand{\shao}{{Shanghai Astronomical Observatory, Chinese Academy of
		Sciences, Shanghai 200030, China}}
\newcommand{\lab}{Key Laboratory of Radio Astronomy, Chinese Academy of
	Sciences, Nanjing 210008, China}
\newcommand{\nav}{Shanghai Key Laboratory of Space Navigation and Positioning 	
	Techniques, Shanghai 200030, China}	
\newcommand{\dc}{National Basic Public Science Data Center, Beijing 100190,
China}

\definecolor{blue}{rgb}{0.0,0.0,1.0}
\newcommand{\revI}[1]{#1}

\newcommand{\p}{^{\prime}}
\newcommand{\pp}{^{\prime\prime}}
\submitjournal{AJ}

\begin{document}
	
\title{VOLKS2: a transient search and localization pipeline for VLBI
observations}
\author{Lei Liu}
\affiliation{\shao}\affiliation{\lab}\affiliation{\nav}\affiliation{\dc}

\author{Zhijun Xu}
\affiliation{\shao}\affiliation{\lab}

\author{Zhen Yan}
\affiliation{\shao}\affiliation{\lab}

\author{Weimin Zheng}
\affiliation{\shao}\affiliation{\lab}\affiliation{\nav}\affiliation{\dc}

\author{Yidan Huang}
\affiliation{\shao}\affiliation{\lab}

\author{Zhong Chen}
\affiliation{\shao}\affiliation{\lab}

\correspondingauthor{Lei Liu}
\email{liulei@shao.ac.cn}

\begin{abstract}
We present VOLKS2, the second release of ``VLBI Observation for transient
Localization Keen Searcher''. The pipeline aims at transient search in regular
VLBI observations as well as detection
of single pulses from known sources in
dedicated VLBI observations. The underlying method takes the idea of geodetic
VLBI data processing, including fringe fitting to maximize the signal power and
geodetic VLBI solving for localization. By filtering the candidate signals with
multiple windows within a baseline and by cross matching with multiple
baselines, RFIs are eliminated effectively. Unlike the station auto spectrum
based method, RFI flagging is not required in the VOLKS2 pipeline. EVN
observation (EL060) is carried out, so as to verify the pipeline’s detection
efficiency and localization accuracy in the whole FoV. The pipeline is
parallelized with MPI and further accelerated with GPU, so as to exploit the
hardware resources of modern GPU clusters. We can prove that, with proper
optimization, VOLKS2 could achieve comparable performance as auto spectrum
based pipelines. All the code and documents are publicly available, in the hope
that our pipeline is useful for radio transient studies.
\end{abstract}

\keywords{instrumentation: interferometers --- methods: numerical}

\section{Introduction}\label{sec:intro}

The search of radio transients is a hot topic in modern astronomy. Fast radio
burst (FRB), as one of the most mysterious transient events, is first
discovered in 2007 \citep{Lorimer2007}. Since then, more than 300 events have
been recorded in TNS\footnote{Transient Name Server (TNS):
https://www.wis-tns.org}. Recently, CHIME present a catalog that contains 535
FRBs between 400 and 800 MHz from
Jul. 25, 2018 to Jul. 1, 2019
\citep{CHIME21}.
 FRB 20200428, discovered by CHIME \citep{CHIME20} and
STARE2 \citep{Bochenek2020}, is assumed as the first event detected inside the
Milky Way. The simultaneous detection in high energy band \citep{INTEGRAL,
HXMT} and the accurate localization by FAST \citep{FAST} confirm the burst
is from the known Galactic magnetar SGR 1935$+$2154, which strongly supports
the magnetar origin of FRB \citep{Yuan2020}.

Although the mainstream of transient search is aperture arrays, e.g., UTMOST
\citep{UTMOST},  CHIME \citep{CHIME18}, LOFAR \citep{LOFAR}, and the popular
search pipelines are station auto spectrum based \citep[PRESTO,][]{PRESTO},
we think it is still worthwhile to develop a pipeline that is based on cross
spectrum and dedicated to VLBI \revI{(Very Long Baseline Interferometry)}  observations. For one thing, transient search
with unknown position could be conducted as commensal task in regular VLBI
observations. E.g., V-FASTR \citep{VFASTR}, LOCATe \citep{LOCATe}.
 Moreover, for those repeating
sources, e.g. repeating FRB, magnetas, pulsars, etc, dedicated VLBI
observations are usually required for following monitoring and high accuracy
localization \citep{Chatterjee2017, Marcote2017}. Therefore, a pipeline that
is able to detect single pulse and carry out localization is very necessary
for VLBI observations.

\citet{L18a} propose a cross spectrum based single pulse search method. It
takes the idea of geodetic fringe fitting, which maximizes the single pulse
signal by fully utilizing the cross spectrum fringe phase information. Once the
single pulse is detected, its
position can be derived immediately with geodetic VLBI solving
\citep{L19}.
Cross spectrum based search method outperforms auto spectrum method by
its ability to extract single pulse signal from highly RFI contaminated data
\citep{L18b}. Moreover, as the concept of ``Satellite Internet Access'' becomes
popular, a large amount of satellites are sent to the Earth orbit.
E.g., the ``Starlink'' project initiated by SpaceX. Those satellites not only
transmit radio signals at a wide range of frequencies, but also reflect FM
broadcastings from the ground. Recent studies suggest those artificial
signals are huge threats for radio astronomy especially transient studies: it
is almost
impossible for station auto spectrum based methods to differentiate all
of these signals. For the detected candidates, cross matchings with NORAD
database are even required for further confirmation \citep{Clery2020,
Prabu2020}. In contrast, RFI contamination is not a big problem for cross
spectrum based method, since the baselines are usually
thousand kilometers long, the probability that RFIs are simultaneously detected
by multiple stations and aligned in frequency domains is extremely small.
Moreover, RFIs could be eliminated
effectively by filtering the candidate signals with multiple windows inside a
baseline and by cross matching with multiple baselines.

In \citet{L19}, we present the first release of VOLKS pipeline that
implements
the cross spectrum based method. However, the pipeline is still in its very
preliminary stage. E.g., no parallel support, no hardware acceleration,
dependence of external software (AIPS\footnote{http://www.aips.nrao.edu}), etc.
What's more, the pipeline is only
tested with CVN \citep[Chinese VLBI Network, ][]{CVN} observation psrf02, in
which
the target is always in the FoV
(Field of View)
center. However, in the actual transient search, the event
could appear anywhere in the FoV.
To verify the pipeline's performance of single pulse search and localization in
the whole FoV, we submit our proposal to EVN \revI{(European VLBI Network)}, in which the target pulsar is
placed in the FoV with different offsets to the FoV center.
This observation is approved and conducted in March 11, 2019.

In this paper, we introduce the second release of the pipeline: VOLKS2 \revI{\citep{volks2}}.
Compared with its first release, it implements many features that have
been verified in the actual data processing. We can demonstrate that, this
release is more suitable for productive deployment.

This paper is organized as follows: Sec.~\ref{sec:pipeline} introduces the
VOLKS2 pipeline, including its algorithm and performance improvement.
In Sec~\ref{sec:obs}, the pipeline is applied to EVN observation EL060. The
single pulse
search and localization result is presented. Sec.~\ref{sec:conclusion} presents
the conclusion.

\section{The VOLKS2 pipeline}\label{sec:pipeline}
\begin{figure}
\plotone{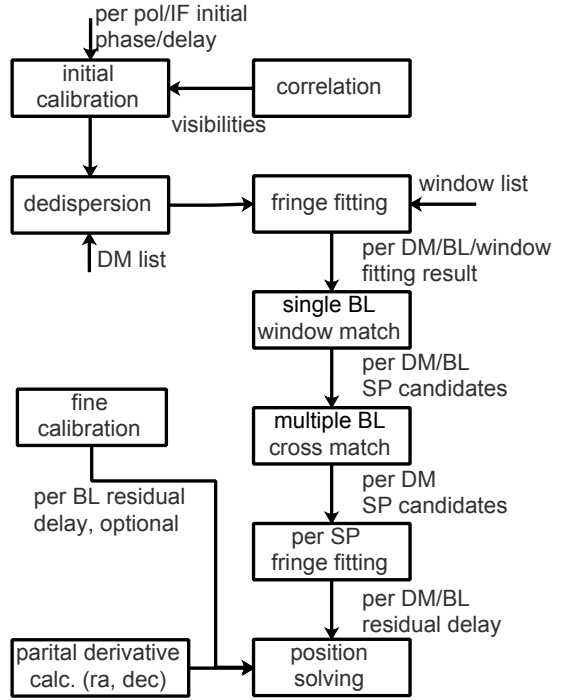}
\caption{Data flow of VOLKS2 pipeline. Visibilities are first loaded into
memory for all baselines (BLs) in a given time range.
Then initial calibration is carried out for every polarization \revI{and IF (Intermediate Frequency)}. Then
visibilities from multiple polarizations are combined together. Dedispersion is carried out for every DM \revI{(Dispersion Measure)}.
After fringe fitting, single baseline window matching and multiple baselines
cross matching, a single pulse \revI{(SP)} list is obtained for each DM. For every single pulse, residual delay of each baseline is derived via fringe fitting. Finally the offset to a priori position is derived with geodetic VLBI solving.	
\label{fig:dataflow}}
\end{figure}

The detailed introduction to the VOLKS2 pipeline has been given in
\citet{L18a, L19}. Significant improvements have been made since the first
release of the pipeline. Fig.~\ref{fig:dataflow} present the data flow of
current pipeline. In this section, we will focus on the updates of the
pipeline and the underlying algorithms.

\subsection{Calibration}
The calibration is divided into two parts, namely initial calibration and fine
calibration.

\subsubsection{Initial calibration}
\begin{figure*}
	\plotone{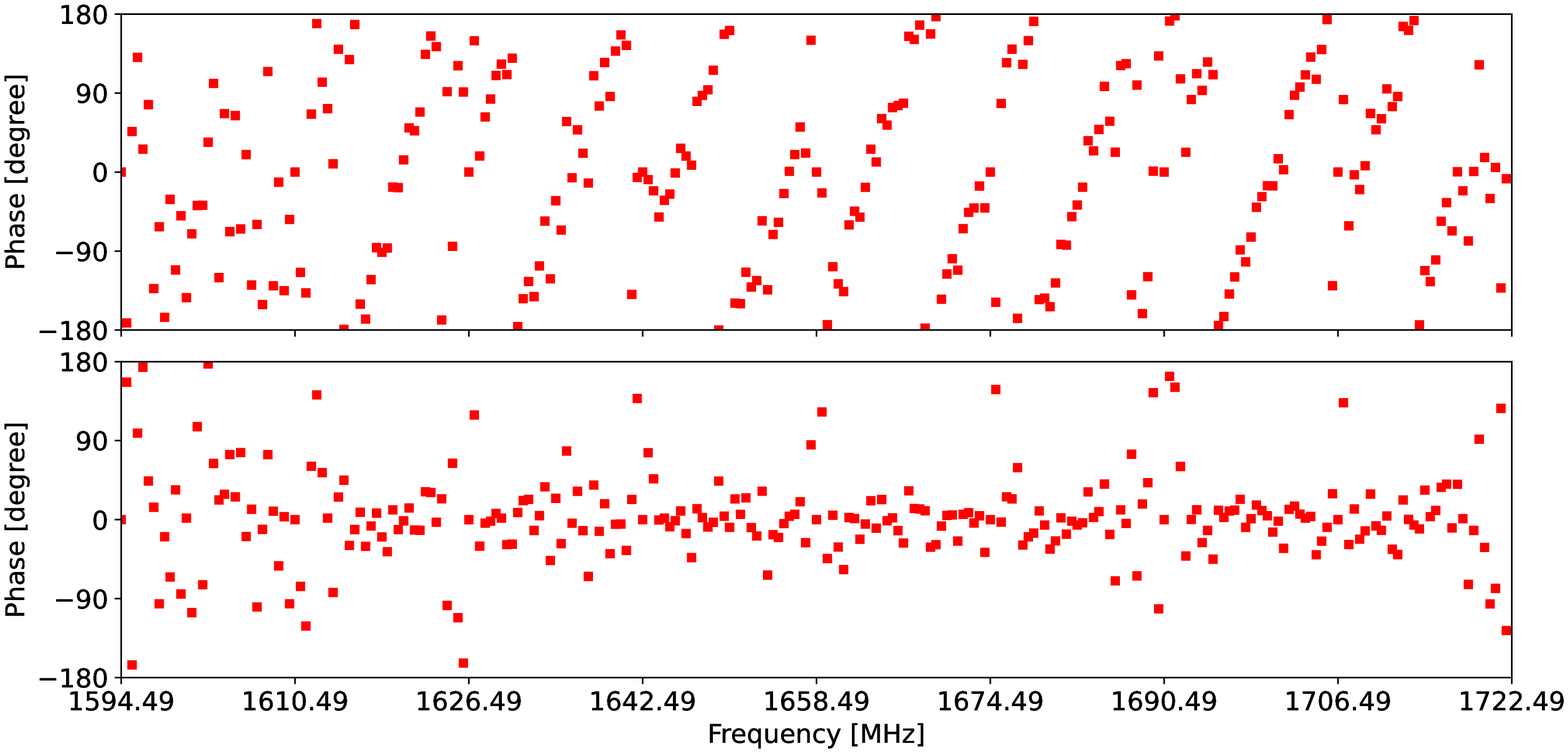}
	\caption{Demonstration of initial calibration. Upper and lower panels
		demonstrate the fringe phase before and after calibration.
		The initial phase and delay are derived for every individual
		polarization, baseline and IF.
		Parameters: EL060 (EVN), J1800$+$3848, Scan 36, LL
		polarization, Ir-Mc baseline, 30~s integration. \label{fig:init_cal}}
\end{figure*}

The purpose of initial calibration is to correct the initial delay
$\Delta\tau_0^n$ and phase $\Delta\phi_0^n$ for every IF ($n$ for IF
index), baseline and polarization. The calibration values (delay and phase) are
derived by carrying out fringe fitting on the strong calibration source. We
assume these quantities reflect the status of the VLBI system and are kept
unchanged across the  observation. Fig.~\ref{fig:init_cal} demonstrates the
fringe phase
before and after initial calibration for the calibration source itself. The
large IF delay is mainly due to station clock offsets. The discrepancies of
initial delay and phase between IFs are small, however still exist. After
calibration, fringe phase become flat and aligned across IFs.

\subsubsection{Fine calibration}
\begin{figure*}
	\plotone{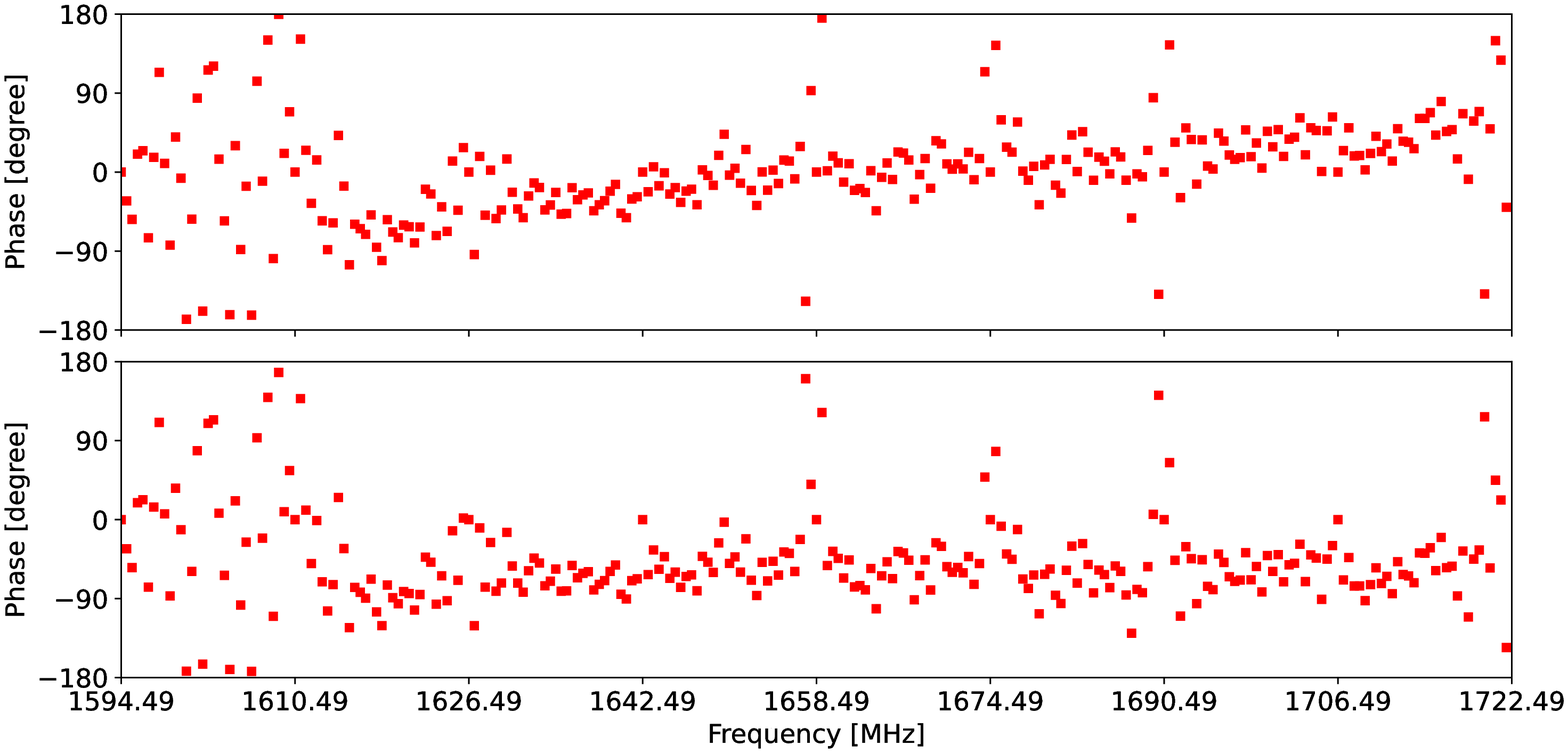}
	\caption{Demonstration of fine calibration. Upper and lower panels
		demonstrate the fringe phase before and after calibration.
		The residual delay is derived by carrying out fringe fitting across
		all IFs.
		Parameters: EL060 (EVN), J0346$+$5434, Scan 37, Ir-Mc
		baseline, 60~s integration.
		\label{fig:fine_cal}}
\end{figure*}

Fine calibration is carried out with reference sources close to the target. We
assume the position of the reference source is accurate. Therefore the non-zero
residual delay across all IFs (upper panel of Fig.~\ref{fig:fine_cal}) is
caused by instruments and atmosphere.  For each
baseline, a residual delay is fitted. The fitting scheme is identical to the
target source as described in Sec.~\ref{sec:fitting}. The fitted value will be
used to correct the baseline residual delay. Unlike phase reference
calibration which requires delay, delay rate and phase information, fine
calibration uses only delay information. This is demonstrated in
Eq.~\ref{eq:solve} of Sec~\ref{sec:loc}.

\subsection{Fringe fitting}\label{sec:fitting}

\begin{figure*}
	\plotone{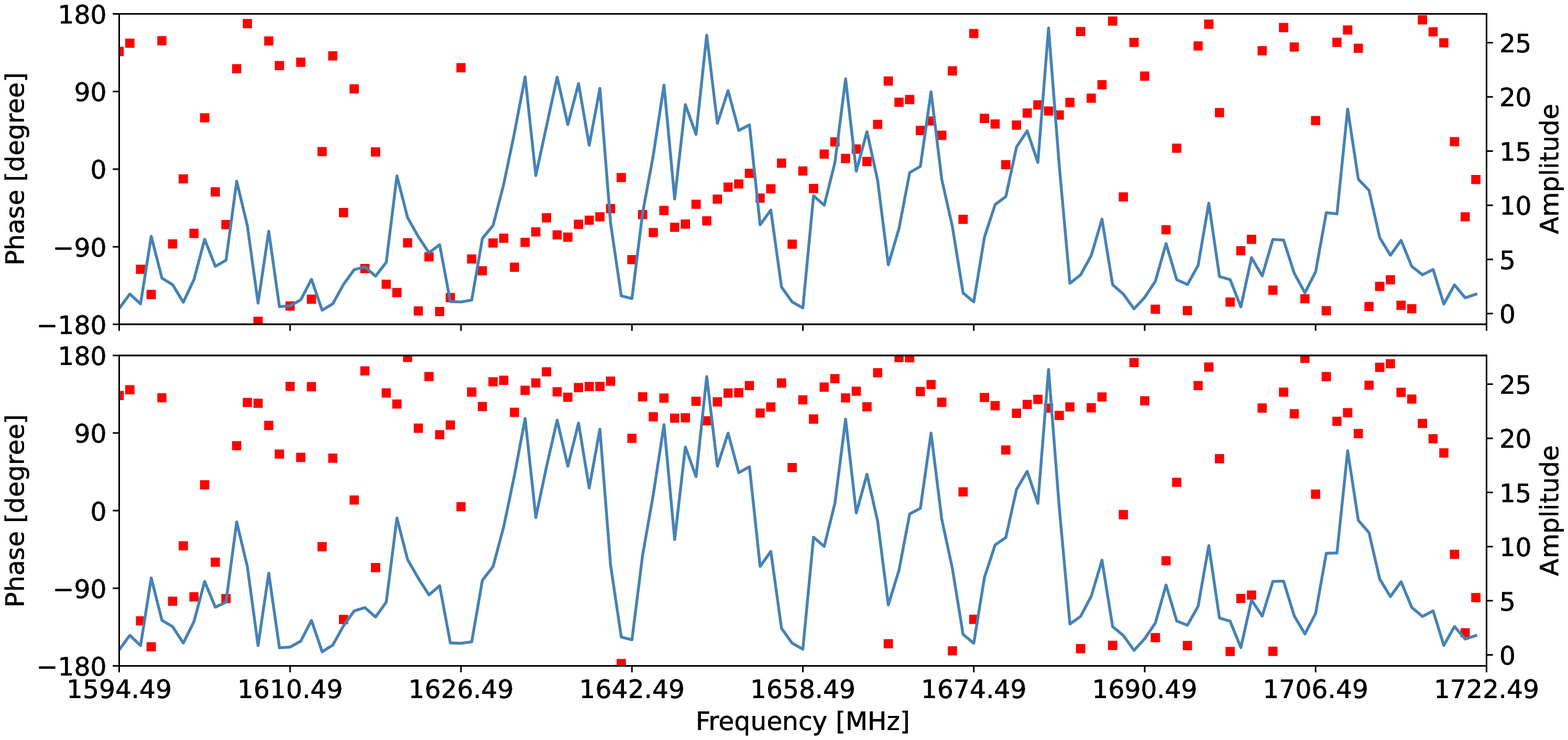}
	\caption{The cross spectrum of a single pulse. The fringe phase
	\revI{(red squares)} and amplitude \revI{(blue lines)} before and after
	fringe fitting are demonstrated in the upper and lower panels,
	respectively.
	For clarity, the complex visibilities of every 64 frequency points are summed coherently. Parameters:
	EL060 (EVN), PSR J0332$+$5434, Scan 50, 2019-03-11 07:52:59.027
	UTC, Ir-Mc baseline, 4.096~ms integration.
	\label{fig:sp}}
\end{figure*}

Fig.~\ref{fig:sp} presents the cross spectrum (fringe phase and
amplitude) of a single pulse detected in the pulsar observation.
\revI{Following the Mk4 format adopted in the geodetic VLBI post processing package HOPS\footnote{https://www.haystack.mit.edu/haystack-observatory-postprocessing-system-hops}, the
	amplitude of each frequency point is normalized
	with the auto spectrum dispersion of the two stations:
	$|\widetilde{s_{ij}}|=|s_{ij}|/(\bar{a_i^2}\bar{a_j^2})^{1/4}$.}
We would like to demonstrate that the fringe fitting process is
effective in extracting single pulses from background noise. In this example,
the fringe phase is clearly detected with just 4 milliseconds integration. A
detailed explanation of the procedure is presented in \citet{L18a}. For
the target single pulse search scan, the length of APs\footnote{\revI{AP is
short for accumulation period. In VLBI data processing it refers to
visibilities within one integration period and IF.}} are in the level of
milliseconds. E.g., 1.024 ms for EL060. For each baseline, several APs are
summed together along the time axis to construct time segments with given
window sizes. Fringe fittings are carried out for
those time segments in each window.
\revI{Window sizes are set to be comparable with the pulsar width. E.g., for
PSR J0332+5434, pulsar width is around 7~ms. The corresponding window sizes are
2~ms, 4~ms, 8~ms, 16~ms. A single pulse is regarded as detected if it appears
in at least 2 windows.}

One thing we have to emphasis is the slightly different fringe fitting scheme.
In \citet{L18a}, standard geodetic VLBI fringe fitting is used, which involves
searching two quantities namely single-band and multi-band delay.
\revI{However this scheme is designed for legacy system, in which the bandwidth
of each IF is much narrower than IF intervals. In the typical astrophysical
observation and in the next generation geodetic VLBI system
\citep[VGOS,][]{VGOS}, IFs are
much wider and always continuous. After calibration,
it is more suitable to treat them as one fringe. Therefore, in} VOLKS2, we search only one delay $\Delta\tau$ across all IFs inside the given band:

\begin{eqnarray}
G(\Delta\tau) = \sum_{n = 0}^{N-1} \sum_{j=1}^{J-1}
S(n, j) e^{-i\Phi(n, j)} \nonumber
\end{eqnarray}
with
\begin{eqnarray}\label{eq:fs}
\Phi(n, j) & = & 2\pi (f_0^n + f_j - f_\mathrm{ref})\Delta\tau\nonumber \\
		 & + & 2\pi f_j\Delta\tau_0^n + \Delta\phi_0^n,
\end{eqnarray}
where $G(\Delta\tau)$ represents the searching function of the corresponding
time segment $S$, $f_j$ is the $j$th frequency point in the $n$th IF,
$f_\mathrm{ref}$ is the reference frequency for fringe fitting,
$\Delta\tau_0^n$ and $\Delta\phi_0^n$ are the initial delay and phase
calibration values of the corresponding IF.

In the actual implementation, delay search is carried out with Fast
Fourier
Transform (FFT). To reduce the effect of cyclic convolution, padding is
required. We choose a padding factor of 4 in the current
pipeline. Standard geodetic fitting scheme conducts 2-D FFT, which
requires a memory
size of at least 16 times (4$\times$4) that of the actual data. VOLKS2
conducts 1-D FFT, which reduces memory consumption by a factor of 4.
This is especially important for GPU \revI(Graphics Processing Unit)
computing: compared with CPU, the size of
GPU memory is quite limited. Lower memory consumption means we can initiate
more computing context within one card, and therefore achieve higher
performance. Moreover, current 1-D scheme provides much higher delay search
resolution, which helps enhancing the SNR of the detected signal.

\subsection{Localization}\label{sec:loc}
Localization is an important step in single pulse search. For VLBI observation,
radio
imaging method is usually adopted, such as the famous ``\textit{realfast}''
pipeline
\citep{realfast} that plays an important role in the first successful
localization of repeating burst FRB121102.
However, such kind of method carries out single pulse search in every snapshot,
which is time
consuming. As pointed out by \citet{L19}, by assuming the single pulse is a
point source,
it is possible to derive the single pulse
position given the relation between the residual delay and the offset to a
priori position. To improve the localization result, the single pulse fringe of
each
baseline must be calibrated. In the first release of VOLKS, this is conducted
in AIPS. The pipeline at that time provides tools for preparing AIPS
inputs and output. The calibration is conducted manually in AIPS.

For the design of VOLKS2, we want to get rid of the dependence of external
software, and make the calibration and localization process simple and
automatic. Therefore, we propose to conduct fine calibration, in which the
residual delay is calibrated with the fringe fitting result of nearby reference
sources:

\begin{equation}\label{eq:solve}
\tau - \tau_\mathrm{cal} = \frac{\partial\tau}{\partial\alpha}\Delta\alpha
+ \frac{\partial\tau}{\partial\delta}\Delta\delta,
\end{equation}
where $\tau$ and $\tau_\mathrm{cal}$ are the single pulse residual delay and
the fine calibration delay for this baseline,
$\frac{\partial\tau}{\partial\alpha}$ and
$\frac{\partial\tau}{\partial\delta}$ are partial derivatives of delay by
Ra and Dec at a priori position, $\Delta\alpha$ and $\Delta\delta$ are
offsets to a priori position. The derived position is the a priori position
corrected by offsets.

According to above equation, by assuming the single pulse as a point source and
taking the idea of geodetic VLBI solving, we may derive the single pulse
position with delay information only. This is different from phase reference
calibration carried out by
AIPS, which uses all information including delay, rate and phase.
Compared with Eq.~\ref{eq:solve}, phase reference calibration might be more
accurate. However, it is more complicated and therefore not suitable for
automatic implementation. Moreover, VOLKS2 pipeline is designed for regular
VLBI observations, phase reference calibration requires the reference source
close to the target source, which is not always available.

\subsection{Performance improvement}\label{subsec:performance}
The purpose of the VOLKS2 pipeline is to carry out single pulse search in VLBI
observations. The designed running
platform is modern GPU cluster. As the most time consuming part, fringe fitting
is our concern for speedup. This part is both data and computationally
intensive. First of all, loading data into memory is already a huge challenge:
in EL060 which is used for single pulse search testing, the
data size of a target search scan is around 650~GB (1.024~ms AP, 190~s
duration, 7 stations, 21 baselines). In contrast, the memory size of one
computing node is in the order of 100~GB. Even if using RAID to improve
disk IO,
the disk reading speed is still no higher than 1~GB/s. Moreover, the fitting
process itself is time consuming: in the above scan, for one DM, we have to
carry out 4 Million 32768 sized FFT and comparable number of matrix sum and
finding maximum amplitude operations.
To make the time consumption of fringe fitting acceptable, in the development
stage, a lot
of efforts have been made to improve the performance and to fully exploit
the hardware resources, such as parallelization and code optimization.

\subsubsection{Parallelization}
The fringe fitting part of the pipeline is fully parallelized with MPI
\revI{(Message Passing Interface)}. When
the program starts, one assignment process and multiple calculation processes
are
initiated. The number of latter one depends on the CPU core number. For each
scan, when a calculation process is idle, it sends task request to
assignment process and obtains the time range. Then it loads the corresponding
data into memory, carries out fringe fitting for all baselines in this time
range, and saves the results in disk. This scheme is easy to understand and
implement. No data communication is required between calculation processes,
which keeps the architecture simple: the MPI part and the fringe fitting part
are fully decoupled. When some processes are loading data, others carry out
calculation. In this way the overhead of disk reading is reduced.

\subsubsection{Code optimization}
\begin{figure}
	\plotone{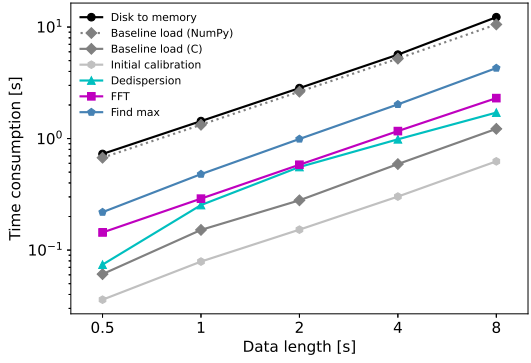}
	\caption{Benchmark results of fringe fitting part before GPU acceleration.
	For
	baseline load (step	2), both NumPy and C results are presented (dashed and
	solid lines). Other steps are conducted with NumPy. \label{fig:bm_numpy}}
\end{figure}

The fringe fitting part could be divided into 6 steps:
\begin{itemize}
	\item[1.] Disk to memory. Load visibilities of give time range to memory.
	\item[2.] Baseline load. Load the required baseline data to the buffer
	array.
	\item[3.] Initial calibration. Carry out IF initial phase and delay
	calibration for buffer data.
	\item[4.] Dedispersion. For buffer data with given DM value.
	\item[5.] FFT. For each AP, carry out FFT along the frequency axis.
	\item[6.] Find max. For each window size, combining the FFTed APs along
	the time axis accordingly; find out the maximum amplitude along the delay
	axis (frequency axis after FFT) for each combined AP.
\end{itemize}
In the above steps, step 1 is conducted once for a given time range; step 2 - 6
are repeated for every baseline; step 4-6 are repeated for every DM value and
baseline. To investigate the time consumption of each step, we carry out
benchmarks. Fig.~\ref{fig:bm_numpy} demonstrates the result before GPU
acceleration. All benchmarks in this work are carried out in a GPU server, of
which the basic configuration is: Intel Xeon 5220R $\times$ 2 (48 cores in
total),
48~TB data storage (RAID 4), 512~GB DDR4 memory, NVIDIA RTX 3080 $\times$ 1,
RTX 2080Ti $\times$ 2.

\begin{figure}
	\plotone{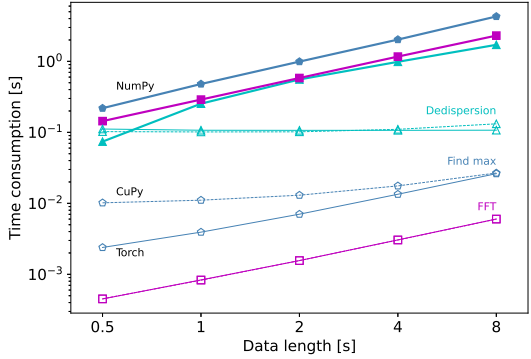}
	\caption{Benchmark results with different frameworks: NumPy for CPU,
	PyTorch and CuPy for GPU. Thick solid, thin solid and thin dashed lines
	correspond to NumPy, PyTorch and CuPy, respectively. Dedispersion, find max
	and FFT steps are plotted with cyan, blue and magenta colors. The line
	styles and colors are consistent with
	Fig.~\ref{fig:bm_numpy}.\label{fig:bm_compare}}
\end{figure}

In a computer system, disk IO is usually slow and is therefore the bottleneck.
In DiFX \citep{DiFX}, visibilities are stored in disk as records. To load the
records of
specific time range to memory, one has to find out the
corresponding start and end indexes. Due to the high latency
of each disk read request, it is not possible to go through all records in the
disk sequentially. Noticing that visibility records are stored in
chronological
order, we implement a binary search routine, which is able to locate
the target record with $\sim\log_2 N$ read requests ($N$ is the total number
of records). This routine greatly reduces the records search time.
However, the disk reading speed is still determined by the hardware
configuration. As a result, the time consumption of step 1 is large and cannot
be easily reduced. Fortunately this step is conducted only once. When averaged
to every individual baseline and DM value, the portion is small.

After loading the records to memory, for each record, one has to parse the
record header, find out the necessary information (time, polarization, baseline
number, frequency), then copy the data to the right place in the buffer array.
This is summarized as step 2, which involves a large number of conditional jump
and memory read/write operations, and is usually not suitable for
script languages such as
Python. Therefore, we rewrite this part in C, compile it to dynamic loading
library and call it in Python with ``ctypes''. With this optimization,
the time consumption of this step is reduced by an order (dotted and solid gray
line in the figure), which makes the following optimization worthwhile.

After rewriting step 2 in C, the actual fringe fitting calculation, step 4 - 6
becomes the bottleneck. We use GPU to achieve further speedup. Two GPU
computing frameworks, Torch and CuPy are used to call GPU resources. The
benchmark is carried out with the NVIDIA RTX 3080 in the server. The result
is demonstrated in Fig.~\ref{fig:bm_compare}. With the help of GPU, step 5 and
6 (FFT and Find max) achieve more than 2 orders speedup. The dedispersion
step (step 4) is also accelerated when $t_\mathrm{seg}\ge 1$~s. However, the
time consumption of this step does not change with data length.
The reason is, for each frequency point, dedispersion invokes the GPU kernel
once and applies a specific time shift to the data. In EL060 which is used for
benchmark, the GPU kernel is invoked 8192 times. Since the time shift operation
is simple, the kernel running time is negligible. However the accumulated
kernel launch overhead, which is independent of data length, becomes
significant.
This part will be further optimized with the update of the pipeline.

One important parameter for optimization is the overall speedup. According to
the analysis above, one may realize that the value depends on the specific
application scenarios: for a typical configuration of EL060 (data length 4~s,
21 baselines), with known DM, the speedup is 7.9~X (209.1~s vs. 26.6~s).
However, if DM search is required, in the case of 20 DM values, the speedup is
23.7~X (1876.1~s vs. 79.0~s). We would like to point out that some
steps, e.g., initial calibration and dedispersion, can be further optimized, so
as to achieve even higher speedup.

\section{Single pulse detection in EL060}\label{sec:obs}
We carry out single pulse detection with EL060 (EVN). The purpose
of the observation is to verify the cross spectrum based single pulse
detection
scheme and the corresponding VOLKS2 pipeline.  Main parameters of the
observation are listed in Tab.~\ref{tab:el060}. The observation is divided
into pulsar part and \revI{RRAT \citep[Rotating Radio Transient, ][]{RRAT}} part. In general, single pulses are
successfully detected and
localized in the pulsar observation. However, no single pulse is detected in
the RRAT
observation. We will present detailed result and analysis in this section.

\begin{deluxetable}{ll}
	\caption{Main parameters of EVN observation EL060. \label{tab:el060}}
	\tablewidth{0pt}
	\tablehead{
		\colhead{Parameter} & \colhead{Setting}
	}
	\startdata
	Experiment code		&	EL060 \\
	Observation date	&	Mar. 11, 2019  (MJD 58553) \\
	Observation time	&	Start:~05h00m00s~UTC\\
						&	Stop:~~08h00m00s~UTC\\
	Station\tablenotemark{a}&	Wb, Ef, Mc, O8, Tr, Hh, \\
						& 	Sv, Zc, Bd, Sr, Ir \\
	Target 				&	RRAT J1819$-$1458, RRAT J1854$+$0306 \\
						&	PSR J0332$+$5434 \\
	Initial calibration	&	J1800$+$3848 \\
	Fine calibration	&	J1825$-$1718, J1907$+$0127, J0346$+$5400 \\
	Polarization 		&	L, R \\
	Frequency			&	1594.49~MHz - 1722.49~MHz, 8 IFs \\
	Bandwidth			&	16~MHz per IF \\
	Sample bit			&	2 \\
	\enddata
	\tablenotetext{a}{Hh, Wb and Ef took part in RRAT observations only. }
\end{deluxetable}

\subsection{Pulsar observation}

This part observes pulsar J0332$+$5434 (B0329$+$54). By placing the pulsar in
the FoV with different offsets to the FoV center, we could investigate the
detection efficiency and the localization accuracy in the whole FoV.
The size (diameter) of
the FoV is estimated as $\sim\lambda/D$. $\lambda$ is set to 18~cm (L
band) according to the observation. $D$ is the antenna diameter, a value of
32~m is selected, which is common for EVN stations (Mc, Tr, Ir, Sv, Zc,
Bd). Above yield a FoV radius of \revI{$r$~=~9.7 arcmin}. When preparing for the
observation schedule, the pulsar is placed at 4 different places in the
FoV by adjusting the offset between the pulsar and the antenna
pointing center. The details of point centers together with the
detection results at that place are
listed in Tab.~\ref{tab:psr_rate}. Among those 4 pointing centers, P0
points to pulsar's a priori position
(Ra $03^\mathrm{h}32^\mathrm{m}59^\mathrm{s}.368$, Dec $54^\circ34\p43\pp.57$).
OF1, OF2 and OF3 point to 3 different offsets by adjusting
declination only (right ascension is kept unchanged).

\subsubsection{Detection rate}\label{sec:psr_rate}

To evaluate the detection rate of each source under the same configuration, we
exclude the Onsala (``O8'') station, since it is not available in the
observation of OF2 and OF3. Moreover, for each scan, the on source time of each
telescope is different. Therefore, we only include single pulses in the time
range that all the considered stations are available for observation.

In total, we have 7 stations (Bd, Ir, Mc, Sr, Sv, Tr, Zc) that consist of 21
baselines. For each baseline, several APs are summed together along the time
axis to construct time segments of different window sizes. We choose window
sizes of 2, 4, 8, 16 APs. A single pulse is assumed to be detected when it
fulfills the follow criterion:
\begin{itemize}
\item For the time segments series of given window size, a single pulse is
selected
if its normalized power exceeds the threshold, which is set to 3 in this
observation.
\item For a give baseline, match single pulses from 4 windows. A single pulse
is selected if it appears in at least 2 windows.
\item Cross match single pulses from multiple baselines. A single pulse is
identified if
it is detected with at least 5 baselines.
\end{itemize}
Note that above criterion, together with the selection of windows sizes, could
be adjusted according to observation.

According to Tab.~\ref{tab:psr_rate}, for each target, we calculate the total
observation time and the ``expected'' single pulse number. Once a single
pulse is detected, we could differentiate whether it is from the pulsar based
on its pulsar phase. Real signals from pulsar and RFIs are labeled as
``detected'' and ``invalid'', respectively. We get lower detection rate and
invalid rate when the pulsar is close to the FoV center (P0 and OF1). However
this does not agree with our expectation: central region of the FoV means
higher antenna response, which should yield \revI{higher} detection rate.
\revI{By investigating the SNR as a function of time, the decrease of SNR
in one of the P0 scan is observed in almost all baselines, which suggest
that this is due to the flux variation of the pulsar. The most possible
explanation of this phenomenon is refractive interstellar scintillation,
which is caused by the
small scale density fluctuations and appears as flux density variations in
both time and frequency domains. According to \citet{Wang2008}, the
scintillation time scale of PSR J0332+5434 is 17.1 minutes, which is
consistent with the dwelling time of each pointing center. Therefore, our
postulation is the detection efficiency is roughly constant in the whole
FoV.
The variation of detection efficiency from 70\% to 90\% is mainly caused by
interstellar scintillation.}

\begin{deluxetable*}{lrrrrrr}
\caption{Parameters of 4 pointing centers and the corresponding detection
results. \revI{``Offset'' is in the fraction of FoV radius.} The right
ascension of the 4 point centers is set to
$03^\mathrm{h}32^\mathrm{m}59^\mathrm{s}.368$ and is kept unchanged.
\label{tab:psr_rate}}
\tablehead{\colhead{Pointing center} & \colhead{Offset} &
\colhead{Dec} & \colhead{Data length} & \colhead{Expected} & \colhead{Detected}
& \colhead{Invalid} }
\startdata
P0 & 0.0 &  $54^\circ34\p43\pp.57$  & \revI{340}~s & 476 & 345 (72.5\%) & 1 (0.3\%) \\
OF1 & 0.2 & 		$+1\p56\pp.02$  & \revI{340}~s & 476 & 424 (89.1\%) & 16 (3.6\%)\\
OF2 & 0.5 & 		$+4\p50\pp.06$ 	& \revI{360}~s & 503 & 406 (80.7\%) & 0 (0.0\%) \\
OF3 & 0.9 & 		$+8\p42\pp.11$  & \revI{306}~s & 429 & 388 (90.4\%) & 0 (0.0\%) \\
\enddata
\end{deluxetable*}

\subsubsection{Localization accuracy}\label{sec:psr_loc}
For single pulses detected with above criterion, we carry out
localization using the method described in \revI{\citet{L19}.
As an
improvement of the pipeline, fine calibration is proposed and
introduced in Sec.~\ref{sec:loc}.} When processing
the data, we find that some further refinements are required to give
reasonable localization result. First of all, Bd related baselines
are excluded. In this work, the delay search range and the related
FFT size is configured for \revI{projected} baselines shorter than
3000~km. Bd related baselines exceed this limit. We can still detect
single pulses with these baselines. However, due to the ambiguity of
the delay search range, the fitted residual delay cannot be used for
localization. Second, baselines with lower SNRs are excluded. For
localization, visibilities that contain the single pulse signal are
extracted and fringe fitted again by following the normalization and
fitting procedure used in HOPS.
 Besides a more accurate fitting result, \revI{SNR is derived
 accordingly.}
 We take a threshold of SNR $>$ 6, which is similar as that adopted
in geodetic VLBI solving. For each single pulse, we exclude baselines that do
not fulfill the above two requirements. Those with at least 3 baselines
are used for localization.

We have to point out that one additional data point (OF3, Scan 40,
31.996~s, Mc-Sr baseline), is excluded from position solving. Although its SNR
is high enough, the fitting result is wrong (probably due to RFI). The position
of the corresponding single pulse solved by including this baseline deviates
significantly from that of other single pulses. Actually this feature can be
used to identify the incorrect baseline fitting result:
\begin{itemize}
\item For the given single pulse, carry out solving several times, each time
exclude one baseline;
\item Check the solving result, solution that includes the incorrect
baseline fitting result deviates significantly from other solutions.
\end{itemize}

\begin{figure*}
	\plotone{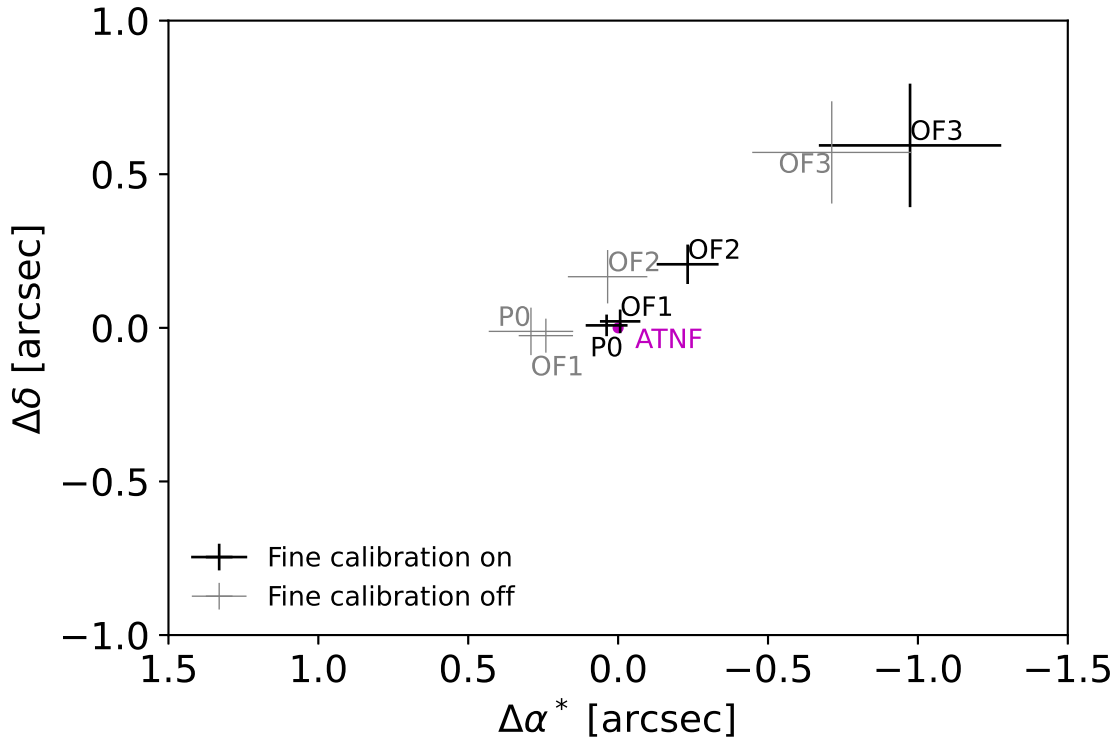}
	\caption{Single pulse Localization result at 4 pointing centers. The center
		of the coordinate (\revI{magenta} dot) corresponds to the position (Ra
		$03^\mathrm{h}32^\mathrm{m}59^\mathrm{s}.418$, Dec
		$54^\circ34\p43\pp.26$) given by the latest (v1.64) ATNF pulsar
		database \citep{ATNF}. The proper motion of the pulsar
		has been taken into account. Black and gray pluses represent the
		localization result with and without fine calibration. The length of
		error bars correspond to the dispersion of localization at given
		pointing center.
		 \label{fig:loc}}
\end{figure*}

Fig.~\ref{fig:loc} demonstrates the localization result. By placing the pulsar
in the FoV with different offsets to the pointing center, we may investigate
the
corresponding localization accuracy. Clearly larger offset yields larger
discrepancy and scatter. This can be explained with the underlying algorithm
of geodetic VLBI solving: the linear relation between the position
offset and the delay is valid only if the target is close enough to a priori
position (small offset). As the offset becomes large, the linear approximation
is not enough to describe its relation with delay, which leads to the large
discrepancy. Meanwhile, scatters are amplified correspondingly. Also note that
in EL060, the offset is always along the declination direction, thus leads to
the systemic trends of discrepancy. In Fig.~\ref{fig:loc}, we present the
localization
result with and without fine calibration. A systematic shift is clearly
observed between the two data sets. With calibration, localization result
derived with P0 as pointing center becomes closer to the reference position,
which is consistent with our analysis: small offset yields small discrepancy.
This also proves that although simple and therefore easy to implement, our fine
calibration scheme is effective in improving the localization accuracy.
In summary, the localization discrepancy is around 1 arcsec when the target
appears at the edge of the FoV, and quickly goes to well within 100~mas at the
FoV center.

\subsection{RRAT observation}\label{sec:rrat}

In this part, we observe RRAT source J1819$-$1458 and J1854$+$0306, so
as to verify the DM search capability of the pipeline. However, in the
80 minutes observation of 2 RRAT sources, no single pulse is detected.
According to RRAT
catalog\footnote{http://astro.phys.wvu.edu/rratalog/}, the flux and width of
the RRAT sources, in particular J1854$+$0306, are quite
similar with that of pulsar J0332$+$5434. It is unreasonable that all single
pulses are missed. Our first suspicion is the newly developed pipeline.
Therefore, we extract the auto spectrum of
each station from the visibility output of the correlator, convert them to the
filterbank format, and carry out single pulse search with STEP. The STEP
transient search
pipeline (Xu et al. 2021, in preparation) is developed in SHAO and tested using
the ASKAP FRB data \citep{ASKAP_DATA}.
In particular, since Ef and Sr are larger than other telescopes, higher
sensitivities are expected. Their data are further inspected by eyes. However,
we are
still not able to detect any single pulse.

By looking up references, one may find that the detected burst rate of RRAT
source strongly depends on telescope sensitivity. E.g., the L-band burst
rates of J1854$+$0306 are 8.9, 84 and 102$\pm$10 per hour for Parkes
\citep{Keane2011}, Arecibo \citep{Deneva2009} and
FAST \citep{Lu2019}, respectively. Moreover, the intensity of single pulse
varies significantly with time. In some epoch, no single pulse is ever detected
\citep{Keane2011}. Note that the size of
Ef is larger than that of Parkes, we may expect a similar or higher detection
rate. \citet{LOCATe} do report the detection of single pulse in the e-EVN data
of J1819$-$1458, which is the first target of EL060. However, in our
observation Ef only observed the second target J1854$+$0306 for 25 minutes.
No single pulse is detected in that short session. One thing we have to admit
is, the possibility that the
failed detection is caused by the VLBI mode observation
still cannot be excluded: in our current treatment, raw data is recorded with
VLBI backend. The single pulse detection is based on the correlated auto
spectrum. Direct recording in pulsar mode might yield better result.

\section{Conclusion}\label{sec:conclusion}


In this work, we introduce the VOLK2 pipeline, which takes the idea of geodetic
VLBI fringe fitting and carries out single \revI{pulse} search in the VLBI cross spectrum.
In the VOLKS2 pipeline, single pulses are identified by fully utilizing the cross
spectrum
fringe phase information. By filtering candidate signals with multiple window
sizes inside one baseline and by cross matching with multiple baselines, RFIs
are eliminated
effectively. Compared with station based single pulse search pipelines, our
pipeline does
not require extra RFI flagging. Once single pulse is detected, its position
could be
derived by geodetic VLBI solving with reasonable accuracy.
The pipeline is designed for the transient search and localization in regular
VLBI observations, as well as single pulse detections with know target
(repeating FRBs, pulsars, etc.) in dedicated VLBI observations.

To verify the pipeline and the underlying cross spectrum based single pulse
search
method, we carry out EVN observation EL060. By placing the target pulsar in the
FoV with different offsets to the FoV center, the detection efficiency
and localization accuracy are investigated.
\revI{We postulate that the detection efficiency is roughly constant in the
whole FoV. The variation could be explained with the flux change caused by
interstellar scintillation.} Moreover, higher localization accuracy is
achieved when the single pulse appears in the FoV center. By utilizing fine
calibration proposed in this work, a localization accuracy of better than
100 mas is achieved. Also
in this observation, no single pulse from RRAT sources is detected. We
postulate this is \revI{because the telescopes are not sensitive enough
to typical RRAT pulses. Therefore} some further
investigation is still required.

To fully utilize the hardware resources of modern GPU clusters, the pipeline
is parallelized with MPI. Moreover, some steps in the fringe fitting part are
accelerated with GPU. Depends on the specific search requirement
and hardware configuration, the pipeline achieves a speedup from 7~X to
23.7~X.
One thing we want to point out is, some unique steps, ``FFT'' and ``Find max'',
of our cross spectrum based pipeline achieves remarkable
speedup
($>$100~X) with the use of GPU, which makes their time consumption
negligible
compared with that of disk read and dedispersion. This suggest that with proper
optimization, the cross spectrum based pipeline could run as fast as
the popular auto spectrum based pipeline.

All the code and document are available in GitHub \revI{repository}
\footnote{https://github.com/liulei/volks2}. We hope the VOLKS2 pipeline
is useful for radio transient studies.

\begin{acknowledgments}
LL would like to thank Dr. Wu Jiang for his kind support in preparing and
processing of the observation data.
This work is supported by the National Natural Science Foundation of China
(Grant Nos. 11903067, 12041301, U2031119, 11973011), CAS Key Technology Talent
Program, Shanghai Leading Talents program, National Basic Public Science Data
Center ``Radio astronomy and deep-space exploration database'' (NBSDC-DB-11),
Natural Science Foundation of Shanghai (Grant No.20ZR1467600). The European
VLBI Network (www.evlbi.org) is a joint facility of independent European,
African, Asian, and North American radio astronomy institutes. Scientific
results from data presented in this publication are derived from the following
EVN project code: EL060.
\end{acknowledgments}

\end{document}